\documentclass[conference,a4paper]{IEEEtran}
\pdfoutput=1 
\usepackage{cite}
\usepackage{amsmath,amssymb,amsfonts}
\usepackage{algorithmic}
\usepackage{graphicx}
\usepackage{textcomp}
\usepackage{xcolor}
\usepackage{empheq}
\usepackage{algorithm}
\usepackage{algorithmic}
\usepackage{subfigure}
\usepackage{stfloats}
\usepackage{verbatim}
\def\BibTeX{{\rm B\kern-.05em{\sc i\kern-.025em b}\kern-.08em
    T\kern-.1667em\lower.7ex\hbox{E}\kern-.125emX}}
\allowdisplaybreaks[4]

\begin{document}

\title{Max-min Rate Deployment Optimization for Backhaul-limited Robotic Aerial 6G Small Cells}

\author{\IEEEauthorblockN{Yuan Liao and Vasilis Friderikos}
\IEEEauthorblockA{Center of Telecommunication Research, 
King's College London,
London, U.K. \\
E-mail: \{yuan.liao, vasilis.friderikos\} @kcl.ac.uk}
}

\maketitle

\begin{abstract}
To overcome the limited on-board battery issue of nominal airborne base stations (ABSs), we are exploring the use of robotic airborne base station (RABS) with energy neutral grasping end-effectors that are able to autonomously perch at tall urban landforms. Specifically, this paper studies a heterogeneous network (HetNet) assisted by a movable RABS as a small cell which connects to a macro base station (MBS) through a limited-capacity wireless backhaul link, which can be deemed as another major challenge. To exploit the potential gains that the mobility of RABS can bring in the system, the minimum rate among all users is maximized by jointly optimizing the RABS deployment, user association and subcarrier allocation. This problem is initially formulated as a binary polynomial optimization (BPO) problem. After reformulating it as a nonconvex quadratically constrained quadratic programming (QCQP), we propose a semidefinite relaxation (SDR) based heuristic method to capture a high-quality solution in polynomial time. Numerical results reveal that deploying a RABS as the small cell can improve the minimum data rate by 95.43\% at most and 33.97\% on average, and the developed SDR heuristic algorithm significantly outperforms the linear relaxation (LR) baseline method.

\end{abstract}


\section{Introduction}
\label{introduction}

Airborne base stations (ABSs) mounted on aerial platforms, such as drones, are envisioned to evolve as an integral part of emerging sixth-generation wireless networks (6G), due to their flexible deployment and controllable 3D mobility.  However, the system performance of ABS-assisted wireless networks is severely restricted by the limited endurance of the on-board battery. To overcome this key shortcoming, robotic airborne base stations (RABSs) with grasping capabilities \cite{friderikos2021airborne} or landing based small cells \cite{landedUAV} have been proposed recently as a mean to prolong the lifetime of ABS-assisted wireless networks. Advances in robotic manipulators \cite{nedungadi2019design} allow the deployment of RABSs that attach autonomously in lampposts (or other tall urban landforms) via energy neutral grasping to act as small cells for multiple hours \cite{friderikos2021airborne}. Therefore, due to their prolonged lifetime, mobility and ease of deployment, RABS might play a significant role in the area of network densification and assist in improving the overall performance of 6G networks.

\begin{figure}[!t]
	\centering
	\includegraphics[width=0.35\textwidth]{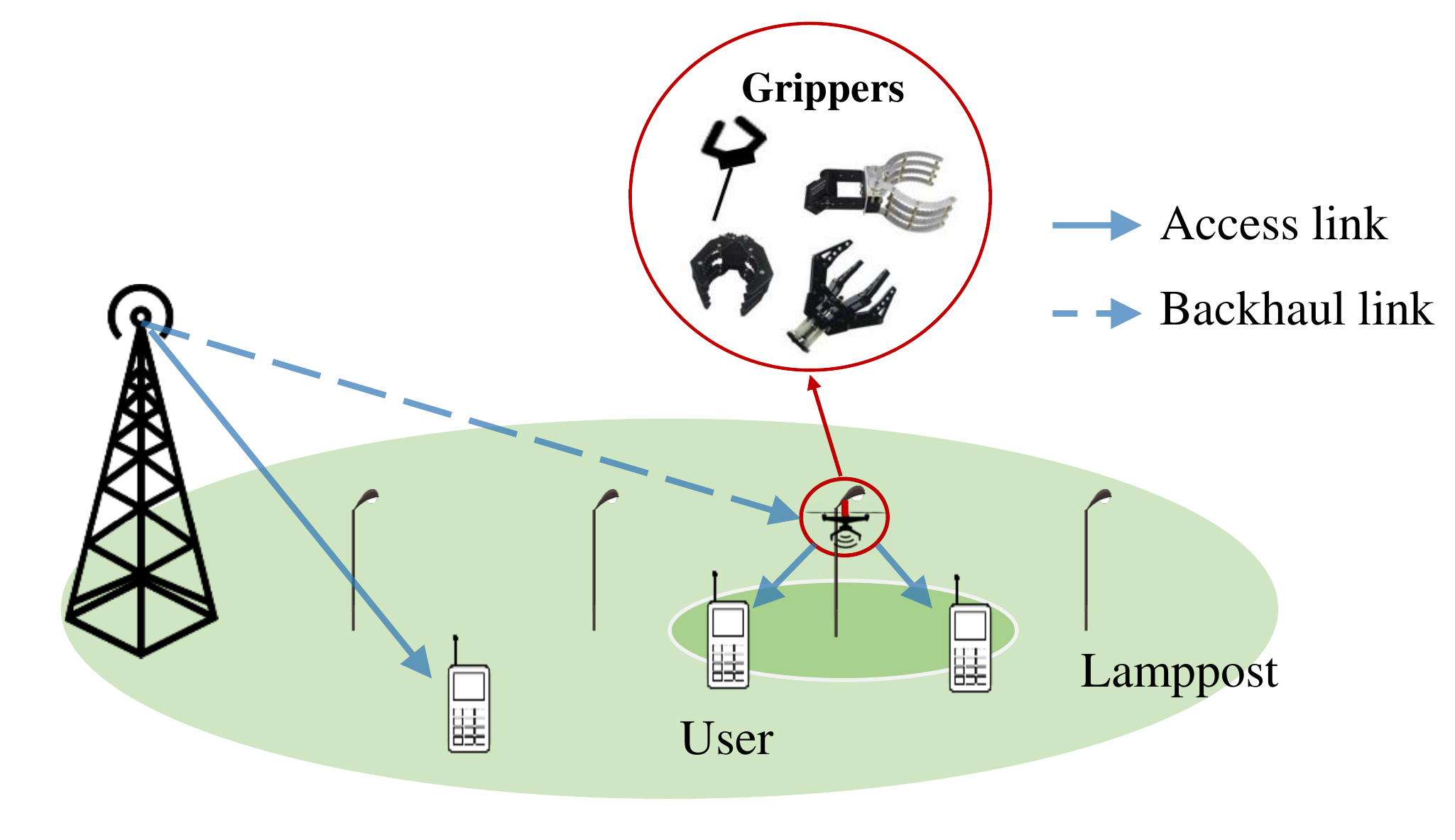}
	\caption{Illustrated use case of RABS-assisted HetNet. }
	\label{figtoy}
\end{figure}

A significant amount of effort has been devoted to overcoming the ABSs endurance issues by developing novel serving protocols and ABS prototypes. The work in \cite{xu2018overcoming} proposes a novel serving scheme that an ABS first offloads the files to a subset of users that cache all the files cooperatively, then each user can receive any file from its nearest neighbor that has cached it via device-to-device communications. Our previous work \cite{friderikos2021airborne} provides a specific introduction of the overall RABS concept, following with aspects related to the optimal operation strategy of the RABS \cite{liao2021optimal}. Furthermore, a RABS-assisted heterogeneous network (HetNet) with mmWave backhauling is studied in \cite{lee2022robotic}. The work in \cite{vaezy2020beamforming} amalgamates beamforming and mmWave communications in ABSs to increase overall network supported quality of service. In \cite{kishk2020aerial}, tethered drones are used to carry ABSs where the tether can transfer both energy and data simultaneously, but their flying range is limited/fixed to the base units located at the ground or a rooftop normally. Also, tethered drones are not flexible to quickly change their locations as is the case for RABS.

On the other hand, unlike terrestrial base stations (BSs) connecting with the core network through a high-capacity fibre link, ABSs require  wireless backhaul due to their frequent movement. Therefore, it is crucial to consider the limited-capacity and position-dependent bachhaul when designing networks with ABSs. In \cite{cicek2020backhaul}, the network profit is maximized by optimizing the ABS location and bandwidth allocation with backhaul consideration. Cache-enabled ABSs are used to support a backhaul-limited network in \cite{kalantari2020wireless}. In \cite{zhang20183}, in-band full-duplex (IBFD) protocol is utilized in an ABS-aided HetNet, in which the access and backhaul channel transmit data on the same spectrum. The following work \cite{zhang2018number} further minimizes the number of required ABSs while incorporating the IBFD technique. 

Motivated by the aforementioned points, this paper studies a HetNet consisting of a MBS and a RABS, which are connected by a limited-capacity wireless hackhaul link, as shown in Fig. \ref{figtoy}. At this point we should stress the difference with nominal ABSs since RABSs are located at lampposts at a height of approximately 5 meters compared to ABSs that hover at much higher altitudes (up to 122 meters according to the current regulation\footnote{www.faa.gov/newsroom/small-unmanned-aircraft-systems-uas-regulations-part-107}). Furthermore, RABS can also rapidly change the lamppost point of attachment and due to energy neutral grasping they can provide service for hours compared to minutes for ABSs. In this paper, with fairness consideration, the minimum rate maximization problem is formulated as a binary polynomial optimization (BPO) problem and then reformulated as a more tractable form, that is, a nonconvex quadratically constrained quadratic programming (QCQP). We then propose a semidefinite relaxation (SDR) based heuristic method to capture an approximated solution in polynomial time. Numerical results show that deploying a RABS can improve the minimum data rate significantly. The performance and efficiency of the proposed SDR heuristic algorithm is discussed as well.

\textbf{Notations:} In this paper, vectors and matrices are
written by boldfaced lower-case and upper-case letters, respectively, e.g., $\mathbf{a}$ and $\mathbf{A}$. $\mathbf{A} \succeq 0$ represents that $\mathbf{A}$ is positive semidefinite. $(\cdot)^T$ denotes the transpose. $\mathrm{Tr}(\cdot)$ calculates the trace. $\mathrm{diag}(\mathbf{A})$ represents a column vector of the diagonal elements of $\mathbf{A}$.

\section{System Model and Problem Formulation}

We consider the downlink case in a HetNet consisting of a macro base station (MBS) and a RABS deployed as a small cell. There are a group of candidate locations distributed in a certain geographical area that can be chosen by RABS for grasping; this set is denoted by $\mathcal{I} = \{1,2,...,I\}$. In practice, the candidate locations can be selected as the roadside lamppost units or other tall urban landforms suitable for grasping and landing \cite{nedungadi2019design}. Denote $\mathcal{J} = \{1,2,...,J\}$ as the user set. Particularly, a user is allowed to associate with at most one BS, but a BS can communicate with multiple users. Besides, we assume that a set of orthogonal subcarriers, denoted by $\mathcal{K} = \{1,2,...,K\}$, are utilized for downlink communication; one user can be provisioned by one or multiple subcarriers while one subcarrier can only be assigned to at most one user to avoid intra-cell interference.

\subsection{Path Loss and Communication Model}
\label{CommunicationModel}

Firstly, considering a user $j$ served by MBS on the subcarrier $k$, the achievable rate can be calculated as,
\begin{equation}
\label{MBSAchRate}
\begin{aligned}
R^{MBS}_{jk} = b_k \cdot \log_2 \big( 1 + p_k h^{MBS}_{j}/\sigma^2 \big) 
\end{aligned} 
\end{equation}
where $b_k$ and $p_k$ are the bandwidth and allocated power of subcarrier $k$, $h^{MBS}_{j}$ is the channel gain between MBS and user $j$, and $\sigma^2$ is the power of background additive white Gaussian noise. Similar as in \cite{han2014enabling,zhang20183,zhang2018number}, to reduce the problem complexity, we assume that the transmit power-spectral density for subcarriers is constant so that $p_k = b_k \cdot \zeta$, where $\zeta$ is the power-spectral density. Similarly, the achievable rate when the user $j$ served by RABS deployed at the candidate location $i$ on the subcarrier $k$ is,
\begin{equation}
\label{RABSAchRate}
\begin{aligned}
R^{RABS}_{ijk} = b_k \cdot \log_2 \big( 1 +  p_k h^{RABS}_{ij}/\sigma^2 \big)
\end{aligned} 
\end{equation}
where $h^{RABS}_{ij}$ is the channel gain between the user $j$ and the RABS placed at the candidate location $i$ acting as a small cell.

As mentioned in section \ref{introduction}, instead of the high-capacity fibre backhaul, the RABS requires a wireless backhaul due to its frequent movement, i.e., grasping at different lampposts. Moreover, wireless backhaul capacity is limited and may also change due to the environmental parameters and the grasping location of RABS \cite{abeta20173gpp}. Accordingly, the backhaul capacity of RABS grasping at candidate location $i$ is denoted as,
\begin{equation}
\label{BackhaulCapacity}
\begin{aligned}
C^{back}_{i} = b^{back} \cdot \log_2 \big( 1 + p^{back} h^{back}_{i}/\sigma^2 \big)
\end{aligned} 
\end{equation}
where $b^{back}$ and $p^{back}$ are the allocated bandwidth and power for the backhaul link, respectively, and $h^{back}_{i}$ is the backhaul channel gain when placing RABS at location $i$.

Furthermore, we adopt the 3GPP path loss model in \cite{abeta20173gpp} for MBS, small cells as well as for the backhaul channel, that is, $128.1+37.6 \log_{10}(d^{MBS}_j)$ and $140.7+36.7 \log_{10}(d^{RABS}_{ij})$ for the path loss in macro and small cells, respectively, where $d^{MBS}_j$ is the distance between MBS and user $j$, and $d^{RABS}_{ij}$ is the distance between the candidate location $i$ and user $j$, both in kilometers. For the wireless backhaul channel, the path losses are $100.7+23.5 \log_{10}(d^{back}_i)$ and $125.2+36.3 \log_{10}(d^{back}_i)$ for line-of-sight (LoS) and non-line-of-sight (NLoS) links, respectively, where $d^{back}_{i}$ denotes the distance between MBS and the candidate location $i$. The LoS probability is calculated by $\min(0.018/d^{back}_i,1)(1- \exp(-d^{back}_i/0.072))+ \exp(-d^{back}_i/0.072)$, shown in the Table A.2.1.1.2-3 of \cite{abeta20173gpp}.

\subsection{Problem Formulation}
\label{problemformulation}

Taking fairness into consideration, we aim to maximize the minimum rate among all users by optimizing the RABS deployment, user association as well as the subcarrier allocation. Thus, there are three sets of binary variables that are used to formulate these decisions. First, $w_{i} \in \{0,1\}$ denotes whether the RABS deploys at the candidate location $i$ ($w_{i} = 1$) or not ($w_{i} = 0$). Second, $x_{j} \in \{0,1\}$ indicates that the user $j$ is served by RABS ($x_{j} = 1$) or MBS ($x_{j} = 0$). Third,  $y_{jk}\in \{0,1\}$ indicates whether the subcarrier $k$ is allocated to the user $j$ ($y_{jk} = 1$) or not ($y_{jk} = 0$). Subsequently, the minimum rate maximization problem can be formulated as,
\begin{subequations}
\begin{align}
\mathrm{(P1):} 
\; & \max_{\{w_{i}\},\{x_{j}\},\atop \{y_{jk}\}} \; \min_{j \in \mathcal{J}} \sum_{i \in \mathcal{I}} \sum_{k \in \mathcal{K}} R^{RABS}_{ijk} w_i x_j y_{jk} \notag \\
\; & \qquad \qquad \quad + \sum_{k \in \mathcal{K}} R^{MBS}_{jk} (1-x_j) y_{jk} \label{Pro1obj} \\
s.t.
\; &  \sum_{i \in \mathcal{I}}w_i \leq 1 \label{Pro1C2}\\
\; &  \sum_{j \in \mathcal{J}}y_{jk} \leq 1, \quad \forall k \in \mathcal{K} \label{Pro1C3}\\
\; &  \sum_{k \in \mathcal{K}}p_k \big( \sum_{j \in \mathcal{J}} x_j y_{jk} \big)  \leq P^{RABS}_{max} \label{Pro1C4}\\
\; &  \sum_{k \in \mathcal{K}}p_k \big( \sum_{j \in \mathcal{J}} (1-x_j) y_{jk} \big) + p^{back} \leq P^{MBS}_{max} \label{Pro1C5} \\
\; & \sum_{j \in \mathcal{J}} \sum_{k \in \mathcal{K}} R^{RABS}_{ijk} w_i x_j y_{jk} \leq C^{back}_{i}, \quad \forall i \in \mathcal{I} \label{Pro1C6} \\
\; & w_{i} \in \{0,1\}, \; x_{j} \in \{0,1\}, \; y_{jk} \in \{0,1\}, \; \forall i,j,k \label{Pro1C7}
\end{align}
\end{subequations}
Specifically, the first term in \eqref{Pro1obj}, i.e. $R^{RABS}_{ijk} w_i x_j y_{jk}$, indicates the achievable rate of user $j$ if it is served by RABS deployed at the candidate location $i$ on subcarrier $k$, and the second term, i.e. $R^{MBS}_{jk} (1-x_j) y_{jk}$, calculates the rate when it is served by MBS on subcarrier $k$. Thus, the objective function \eqref{Pro1obj} indicates the data rate of user $j$ whether it is served by MBS or RABS. Eq. \eqref{Pro1C2} represents that there is at most one RABS can be deployed. Eq. \eqref{Pro1C3} limits that each subcarrier can only be assigned to at most one user to avoid intra-cell interference. Eq. \eqref{Pro1C4} and \eqref{Pro1C5} are the power budget constraints of RABS and MBS, where $P^{MBS}_{max}$ and $P^{RABS}_{max}$ are the maximum transmit power of MBS and RABS, respectively. Eq. \eqref{Pro1C6} represents that the total access data rate served by RABS cannot exceed the backhaul capacity.

It can be seen that (P1) is a BPO problem, which is hard to solve. In the following section \ref{ProReformSDR}, (P1) is first reformulated as a more tractable form, that is, a nonconvex QCQP, and then solved by a proposed SDR heuristic method efficiently.

\section{Problem Reformulation and SDR Heuristic}
\label{ProReformSDR}

To alleviate the difficulties in solving the NP-hard BPO problem, in this section, (P1) is first reformulated as a nonconvex QCQP by introducing additional variables and constraints. We then propose an SDR heuristic algorithm to capture a high-quality solution in polynomial time.

\subsection{QCQP Reformulation}
\label{QCQPReformulation}

The basic idea of reformulating (P1) is to eliminate the cubic terms in \eqref{Pro1obj} and \eqref{Pro1C6} by introducing additional variables and constraints, and replace the binary constraints \eqref{Pro1C7} by introducing a set of quadratic constraints. 

\textit{\textbf{Lemma 1:} (P1) can be refomulated as the following problem (P2) without loss of optimality,}
\begin{subequations}
\begin{align}
\mathrm{(P2):} 
\; & \max_{\{w_{i}\},\{x_{j}\},\atop\{y_{jk}\},\{s_{jk}\}} \; \min_{j \in \mathcal{J}} \; \sum_{i \in \mathcal{I}} \sum_{k \in \mathcal{K}} R^{RABS}_{ijk} w_i s_{jk} \notag \\
\; & \qquad \qquad \quad \; \; + \sum_{k \in \mathcal{K}} R^{MBS}_{jk} (y_{jk}-s_{jk}) \label{Pro2obj} \\
s.t.
\; &  \sum_{i \in \mathcal{I}}w_i \leq 1 \label{Pro2C2} \\
\; &  \sum_{j \in \mathcal{J}}y_{jk} \leq 1, \quad \forall k \in \mathcal{K} \label{Pro2C3}\\
\; &  \sum_{k \in \mathcal{K}} \sum_{j \in \mathcal{J}} p_k s_{jk} \leq P^{RABS}_{max} \label{Pro2C4}\\
\; &  \sum_{k \in \mathcal{K}} \sum_{j \in \mathcal{J}} p_k (y_{jk}-s_{jk}) + p^{back} \leq P^{MBS}_{max} \label{Pro2C5} \\
\; & \sum_{j \in \mathcal{J}} \sum_{k \in \mathcal{K}} R^{RABS}_{ijk} w_i s_{jk} \leq C^{back}_{i}, \quad \forall i \in \mathcal{I} \label{Pro2C6} \\
\; & s_{jk} \leq x_j, \quad \forall j \in \mathcal{J}, \forall k \in \mathcal{K} \label{Pro2C7} \\
\; &  s_{jk} \leq y_{jk}, \quad \forall j \in \mathcal{J}, \forall k \in \mathcal{K} \label{Pro2C8} \\
\; & s_{jk} \geq x_j + y_{jk} - 1, \quad \forall j \in \mathcal{J}, \forall k \in \mathcal{K} \label{Pro2C9} \\
\; & w_{i}(w_{i} - 1) = 0, \quad \forall i \in \mathcal{I} \label{Pro2C10} \\
\; & x_{j}(x_{j} - 1) = 0, \quad \forall j \in \mathcal{J} \label{Pro2C11} \\
\; & y_{jk}(y_{jk} - 1) = 0, \quad \forall j \in \mathcal{J}, \forall k \in \mathcal{K} \label{Pro2C12} \\
\; & s_{jk}(s_{jk} - 1) = 0, \quad \forall j \in \mathcal{J}, \forall k \in \mathcal{K} \label{Pro2C13}
\end{align}
\end{subequations}

\textit{Proof:} The reformulation includes two steps. In the first step, each product term $x_j y_{jk}$ in \eqref{Pro1obj} and \eqref{Pro1C4}-\eqref{Pro1C6} is replaced by a corresponding binary variable $s_{jk}$ with additional constraints \eqref{Pro2C7}- \eqref{Pro2C9}. It can be proven that this step does not lose any optimality because all of $x_j$, $y_{jk}$ and $s_{jk}$ can only take values from $\{0,1\}$. In the second step, each binary constraint is replaced by a corresponding quadratic constraint in \eqref{Pro2C10}-\eqref{Pro2C13}, e.g., the constraint $w_i \in \{0,1\}$ in \eqref{Pro1C7} can be rewritten as $w_i (w_i-1) = 0$ in \eqref{Pro2C10} equally. This completes the proof of Lemma 1. $\square$

For notational convenience, we vectorize the variables and parameters in problem (P2). Define variable vectors as $\mathbf{w} = [w_1,w_2,...,w_I]^T \in \mathbb{R}^{I}$,  $\mathbf{x} = [x_1,x_2,...,x_J]^T \in \mathbb{R}^{J}$,  $\mathbf{y} = [y_1,y_2,...,y_L]^T \in \mathbb{R}^{L}$ and $\mathbf{s} = [s_1,s_2,...,s_L]^T \in \mathbb{R}^{L}$, where $L = J \times K$ is an introduced index number to reshape the sets $\{y_{jk}\}$ and $\{s_{jk}\}$ as vectors. Define a new decision vector containing these variables as $\mathbf{z} = [\mathbf{w}^T, \mathbf{x}^T, \mathbf{y}^T, \mathbf{s}^T]^T \in \mathbb{R}^{I+J+2L}$. (P2) can be then rewritten as, 
\begin{subequations}
\begin{align}
\mathrm{(P3):} 
\; & \max_{\mathbf{z}} \; \min_{j \in \mathcal{J}} \; \mathbf{z}^T \mathbf{Q}^{O}_j \mathbf{z} + (\mathbf{q}^{O}_j)^T \mathbf{z} \label{Pro3obj} \\
s.t.
\; &  (\mathbf{q}^{C_1})^T \mathbf{z} \leq 1 \label{Pro3C2} \\
\; &  (\mathbf{q}^{C_2}_k)^T \mathbf{z} \leq 1, \quad \forall k \in \mathcal{K} \label{Pro3C3}\\
\; &  (\mathbf{q}^{C_3})^T \mathbf{z} \leq P^{RABS}_{max} \label{Pro3C4}\\
\; &  (\mathbf{q}^{C_4})^T \mathbf{z} \leq P^{MBS}_{max} - p^{back}  \label{Pro3C5} \\
\; & \mathbf{z}^T \mathbf{Q}^{C_5}_i \mathbf{z} \leq C^{back}_{i}, \quad \forall i \in \mathcal{I} \label{Pro3C6} \\
\; & (\mathbf{q}^{C_6})^T \mathbf{z}  \leq 0, \quad \forall l \in \mathcal{L} \label{Pro3C7} \\
\; & (\mathbf{q}^{C_7}_{l})^T \mathbf{z}  \leq 0, \quad \forall l \in \mathcal{L} \label{Pro3C8} \\
\; & (\mathbf{q}^{C_8}_{l})^T \mathbf{z}  \leq 1, \quad \forall l \in \mathcal{L} \label{Pro3C9} \\
\; & \mathbf{z}^T \mathbf{Q}^{C_9}_i \mathbf{z} + (\mathbf{q}^{C_9}_i)^T \mathbf{z} = 0, \quad \forall i \in \mathcal{I} \label{Pro3C10} \\
\; & \mathbf{z}^T \mathbf{Q}^{C_{10}}_j \mathbf{z} + (\mathbf{q}^{C_{10}}_j)^T \mathbf{z} = 0, \quad \forall j \in \mathcal{J} \label{Pro3C11} \\
\; & \mathbf{z}^T \mathbf{Q}^{C_{11}}_l \mathbf{z} + (\mathbf{q}^{C_{11}}_l)^T \mathbf{z} = 0, \quad \forall l \in \mathcal{L} \label{Pro3C12} \\
\; & \mathbf{z}^T \mathbf{Q}^{C_{12}}_l \mathbf{z} + (\mathbf{q}^{C_{12}}_l)^T \mathbf{z} = 0, \quad \forall l \in \mathcal{L} \label{Pro3C13}
\end{align}
\end{subequations}where $ \mathcal{L} \triangleq \mathcal{J} \times \mathcal{K}$ is defined as the Cartesian product of two sets $\mathcal{J}$ and $\mathcal{K}$, $\{\mathbf{Q}^{O}_j\}$, $\{\mathbf{q}^{O}_j\}$, $\mathbf{q}^{C_1}$, $\{\mathbf{q}^{C_2}_k\}$, $\mathbf{q}^{C_3}$, $\mathbf{q}^{C_4}$, $\{\mathbf{Q}^{C_5}_i\}$, $\{\mathbf{q}^{C_6}_l\}$, $\{\mathbf{q}^{C_7}_l\}$, $\{\mathbf{q}^{C_8}_l\}$, $\{\mathbf{Q}^{C_9}_i\}$, $\{\mathbf{q}^{C_9}_i\}$, $\{\mathbf{Q}^{C_{10}}_j\}$, $\{\mathbf{q}^{C_{10}}_j\}$, $\{\mathbf{Q}^{C_{11}}_l\}$, $\{\mathbf{q}^{C_{11}}_l\}$, $\{\mathbf{Q}^{C_{12}}_l\}$ and $\{\mathbf{q}^{C_{12}}_l\}$ are the parameter matrices and vectors which are one-to-one corresponding to \eqref{Pro2obj}-\eqref{Pro2C13}, e.g., $\mathbf{q}^{C_1}  = [\mathbf{1}_{1\times I}, \mathbf{0}_{1\times (J+2L)}]^T$ is the vector representation of the parameters in \eqref{Pro2C2}. 
It can be observed that (P3) is a nonconvex QCQP that is NP-hard \cite{luo2010semidefinite}.

\subsection{SDR Heuristic Algorithm}
\label{SDRHeuristic}

SDR is a computationally efficient technique to provide an upper bound of a nonconvex QCQP and has been widely applied in the area of signal processing and communications \cite{luo2010semidefinite,kalantari2020wireless}. In this subsection, we propose an SDR heuristic algorithm for (P3), the basic idea of which is to solve the SDR of (P3) and then construct a feasible solution from the obtained SDR result through a refinement procedure. 

\textit{1) Semidefinite Relaxation for (P3):} To apply the SDR technique to (P3), we would first transform the inhomogeneous QCQP problem (P3) to a homogeneous form by introducing an auxiliary variable. 

\textit{\textbf{Lemma 2:} Introducing an auxiliary variable $t$ and defining a decision vector as $\widetilde{\mathbf{z}} = [\mathbf{z}^T, t]^T \in \mathbb{R}^{I+J+2L+1}$, (P3) can be homogenized as the following (P4) without loss of optimality,
\begin{subequations}
\begin{align}
\mathrm{(P4):} 
\; & \max_{\widetilde{\mathbf{z}}} \; \min_{j \in \mathcal{J}} \; \widetilde{\mathbf{z}}^T \widetilde{\mathbf{Q}}^{O}_j \widetilde{\mathbf{z}} \label{Pro4obj} \\
s.t.
\; &  \widetilde{\mathbf{z}}^T \widetilde{\mathbf{Q}}^{C_1} \widetilde{\mathbf{z}} \leq 1 \label{Pro4C2} \\
\; &  \widetilde{\mathbf{z}}^T \widetilde{\mathbf{Q}}^{C_2}_k \widetilde{\mathbf{z}} \leq 1, \quad \forall k \in \mathcal{K} \label{Pro4C3}\\
\; &  \widetilde{\mathbf{z}}^T \widetilde{\mathbf{Q}}^{C_3} \widetilde{\mathbf{z}} \leq P^{RABS}_{max} \label{Pro4C4}\\
\; &  \widetilde{\mathbf{z}}^T \widetilde{\mathbf{Q}}^{C_4} \widetilde{\mathbf{z}} \leq P^{MBS}_{max} - p^{back}  \label{Pro4C5} \\
\; & \widetilde{\mathbf{z}}^T \widetilde{\mathbf{Q}}^{C_5}_i \widetilde{\mathbf{z}} \leq C^{back}_{i}, \quad \forall i \in \mathcal{I} \label{Pro4C6} \\
\; & \widetilde{\mathbf{z}}^T \widetilde{\mathbf{Q}}^{C_6}_l \widetilde{\mathbf{z}}  \leq 0, \quad \forall l \in \mathcal{L} \label{Pro4C7} \\
\; & \widetilde{\mathbf{z}}^T \widetilde{\mathbf{Q}}^{C_7}_l \widetilde{\mathbf{z}}  \leq 0, \quad \forall l \in \mathcal{L} \label{Pro4C8} \\
\; & \widetilde{\mathbf{z}}^T \widetilde{\mathbf{Q}}^{C_8}_l \widetilde{\mathbf{z}}  \leq 1, \quad \forall l \in \mathcal{L} \label{Pro4C9} \\
\; & \widetilde{\mathbf{z}}^T \widetilde{\mathbf{Q}}^{C_9}_i \widetilde{\mathbf{z}} = 0, \quad \forall i \in \mathcal{I} \label{Pro4C10} \\
\; & \widetilde{\mathbf{z}}^T \widetilde{\mathbf{Q}}^{C_{10}}_j \widetilde{\mathbf{z}} = 0, \quad \forall j \in \mathcal{J} \label{Pro4C11} \\
\; & \widetilde{\mathbf{z}}^T \widetilde{\mathbf{Q}}^{C_{11}}_l \widetilde{\mathbf{z}} = 0, \quad \forall l \in \mathcal{L} \label{Pro4C12} \\
\; & \widetilde{\mathbf{z}}^T \widetilde{\mathbf{Q}}^{C_{12}}_l \widetilde{\mathbf{z}} = 0, \quad \forall l \in \mathcal{L} \label{Pro4C13} \\
\; & \big(\widetilde{\mathbf{z}}_{I+J+2L+1}\big)^2 = 1 \label{Pro4C14}
\end{align}
\end{subequations}
where 
\begin{subequations}
\begin{align*}
& \widetilde{\mathbf{Q}}^{O}_j = 
\begin{bmatrix}
\mathbf{Q}^{O}_j & \frac{1}{2}\mathbf{q}^{O}_j \\
\frac{1}{2}(\mathbf{q}^{O}_j)^T & 0
\end{bmatrix},
\widetilde{\mathbf{Q}}^{C_1} = 
\begin{bmatrix}
\mathbf{0} & \frac{1}{2}\mathbf{q}^{C_1} \\
\frac{1}{2}(\mathbf{q}^{C_1})^T & 0
\end{bmatrix}, \notag \\
& \widetilde{\mathbf{Q}}^{C_2}_k = 
\begin{bmatrix}
\mathbf{0} & \frac{1}{2}\mathbf{q}^{C_2}_k \\
\frac{1}{2}(\mathbf{q}^{C_2}_k)^T & 0
\end{bmatrix},
\widetilde{\mathbf{Q}}^{C_3} = 
\begin{bmatrix}
\mathbf{0} & \frac{1}{2}\mathbf{q}^{C_3} \\
\frac{1}{2}(\mathbf{q}^{C_3})^T & 0
\end{bmatrix}, \notag \\
& \widetilde{\mathbf{Q}}^{C_4} = 
\begin{bmatrix}
\mathbf{0} & \frac{1}{2}\mathbf{q}^{C_4} \\
\frac{1}{2}(\mathbf{q}^{C_4})^T & 0
\end{bmatrix},
\widetilde{\mathbf{Q}}^{C_5}_i = 
\begin{bmatrix}
\mathbf{Q}^{C_5}_i & \mathbf{0} \\
\mathbf{0}^T & 0
\end{bmatrix}, \notag \\
& \widetilde{\mathbf{Q}}^{C_6}_l = 
\begin{bmatrix}
\mathbf{0} & \frac{1}{2}\mathbf{q}^{C_6}_l \\
\frac{1}{2}(\mathbf{q}^{C_6}_l)^T & 0
\end{bmatrix},
\widetilde{\mathbf{Q}}^{C_7}_l = 
\begin{bmatrix}
\mathbf{0} & \frac{1}{2}\mathbf{q}^{C_7}_l \\
\frac{1}{2}(\mathbf{q}^{C_7}_l)^T & 0
\end{bmatrix}, \notag \\
& \widetilde{\mathbf{Q}}^{C_8}_l = 
\begin{bmatrix}
\mathbf{0} & \frac{1}{2}\mathbf{q}^{C_8}_l \\
\frac{1}{2}(\mathbf{q}^{C_8}_l)^T & 0
\end{bmatrix},
\widetilde{\mathbf{Q}}^{C_9}_i = 
\begin{bmatrix}
\mathbf{Q}^{C_9}_i & \frac{1}{2}\mathbf{q}^{C_9}_i \\
\frac{1}{2}(\mathbf{q}^{C_9}_i)^T & 0
\end{bmatrix}, \notag \\
& \widetilde{\mathbf{Q}}^{C_{10}}_j \! = \!
\begin{bmatrix}
\mathbf{Q}^{C_{10}}_j & \frac{1}{2}\mathbf{q}^{C_{10}}_j \\
\frac{1}{2}(\mathbf{q}^{C_{10}}_j)^T & 0
\end{bmatrix},
\widetilde{\mathbf{Q}}^{C_{11}}_l \! = \!
\begin{bmatrix}
\mathbf{Q}^{C_{11}}_l & \frac{1}{2}\mathbf{q}^{C_{11}}_l \\
\frac{1}{2}(\mathbf{q}^{C_{11}}_l)^T & 0
\end{bmatrix}, \notag \\
& \widetilde{\mathbf{Q}}^{C_{12}}_l \! = \!
\begin{bmatrix}
\mathbf{Q}^{C_{12}}_l & \frac{1}{2}\mathbf{q}^{C_{12}}_l \\
\frac{1}{2}(\mathbf{q}^{C_{12}}_l)^T & 0
\end{bmatrix},
\end{align*}
\end{subequations}
And \eqref{Pro4C14} implicates that the introduced auxiliary variable $t$ should satisfy the equation $t^2 = 1$.
}

\textit{Proof:} Please refer to \cite{luo2010semidefinite}. $\square$

To obtain the SDR of (P4), we define a decision matrix as $\mathbf{Z} \triangleq \widetilde{\mathbf{z}} {\widetilde{\mathbf{z}}}^T$. Afterwards, each LHS in \eqref{Pro4C2}-\eqref{Pro4C13} can be rewritten in an equivalent form $\mathrm{Tr}(\widetilde{\mathbf{Q}}\mathbf{Z})$ with a semidefinite constraint $\mathbf{Z} \succeq 0$ and a rank-one constraint, that is, 
\setcounter{equation}{7}
\begin{equation}
\label{rankconstraint}
\begin{aligned}
\mathrm{rank}(\mathbf{Z}) = 1
\end{aligned} 
\end{equation}
By dropping the nonconvex rank-one constraint \eqref{rankconstraint}, we relax problem (P4) into the following (P5),
\begin{subequations}
\begin{align}
\mathrm{(P5):} 
\; & \max_{\mathbf{Z} \succeq 0 } \; \min_{j \in \mathcal{J}} \; \mathrm{Tr}(\widetilde{\mathbf{Q}}^{O}_j\mathbf{Z}) \label{Pro5obj} \\
s.t.
\; &  \mathrm{Tr}(\widetilde{\mathbf{Q}}^{C_1}\mathbf{Z}) \leq 1 \label{Pro5C2} \\
\; &  \mathrm{Tr}(\widetilde{\mathbf{Q}}^{C_2}_k\mathbf{Z}) \leq 1, \quad \forall k \in \mathcal{K} \label{Pro5C3}\\
\; &  \mathrm{Tr}(\widetilde{\mathbf{Q}}^{C_3}\mathbf{Z}) \leq P^{RABS}_{max} \label{Pro5C4}\\
\; &  \mathrm{Tr}(\widetilde{\mathbf{Q}}^{C_4}\mathbf{Z}) \leq P^{MBS}_{max} - p^{back}  \label{Pro5C5} \\
\; & \mathrm{Tr}(\widetilde{\mathbf{Q}}^{C_5}_i\mathbf{Z}) \leq C^{back}_{i}, \quad \forall i \in \mathcal{I} \label{Pro5C6} \\
\; & \mathrm{Tr}(\widetilde{\mathbf{Q}}^{C_6}_l\mathbf{Z})  \leq 0, \quad \forall l \in \mathcal{L} \label{Pro5C7} \\
\; & \mathrm{Tr}(\widetilde{\mathbf{Q}}^{C_7}_l\mathbf{Z})  \leq 0, \quad \forall l \in \mathcal{L} \label{Pro5C8} \\
\; & \mathrm{Tr}(\widetilde{\mathbf{Q}}^{C_8}_l\mathbf{Z})  \leq 1, \quad \forall l \in \mathcal{L} \label{Pro5C9} \\
\; & \mathrm{Tr}(\widetilde{\mathbf{Q}}^{C_9}_i\mathbf{Z}) = 0, \quad \forall i \in \mathcal{I} \label{Pro5C10} \\
\; & \mathrm{Tr}(\widetilde{\mathbf{Q}}^{C_{10}}_j\mathbf{Z}) = 0, \quad \forall j \in \mathcal{J} \label{Pro5C11} \\
\; & \mathrm{Tr}(\widetilde{\mathbf{Q}}^{C_{11}}_l\mathbf{Z}) = 0, \quad \forall l \in \mathcal{L} \label{Pro5C12} \\
\; & \mathrm{Tr}(\widetilde{\mathbf{Q}}^{C_{12}}_l\mathbf{Z}) = 0, \quad \forall l \in \mathcal{L} \label{Pro5C13} \\
\; & \mathbf{Z}_{I+J+2L+1,I+J+2L+1} = 1 \label{Pro5C14} 
\end{align}
\end{subequations}
To avoid the non-differentiable min function in \eqref{Pro5obj}, we introduce an auxiliary variable $\eta \geq 0$ denoting the minimum rate among all users, implicated by a group of additional constraints $\mathrm{Tr}(\widetilde{\mathbf{Q}}^{O}_j\mathbf{Z}) \geq \eta, \; \forall j \! \in \! \mathcal{J}$. This equivalent problem is a semidefinite programming that can be solved in polynomial time by convex optimization tools such as CVX\cite{cvx}.

\textit{2) Construct a feasible solution:} Because the rank-one constraint \eqref{rankconstraint} is relaxed in (P5), the captured solution, denoted by $\mathbf{Z}^{sdr}$, may be not feasible for (P4). A normally used refinement approach is randomization method \cite{luo2010semidefinite}, which is, however, cannot yield a feasible solution for (P4) due to the coupled constraints \eqref{Pro4C2}-\eqref{Pro4C13}. Hereafter, we propose a refinement method to generate a feasible solution for (P4) from $\mathbf{Z}^{sdr}$ based on the randomized rounding technique and greedy search method.

\textit{\textbf{Lemma 3:} If $\mathbf{Z}^{sdr}$ satisfies the rank-one constraint \eqref{rankconstraint}, $\mathbf{z}^{sdr}$ defined by the following \eqref{diagonalZ} is optimal for (P3).
\begin{equation}
\label{diagonalZ}
\begin{aligned}
\mathbf{z}^{sdr} \triangleq [\mathrm{diag}(\mathbf{Z}^{sdr})_1,...,\mathrm{diag}(\mathbf{Z}^{sdr})_{I+J+2L}]^T
\end{aligned} 
\end{equation}}

\textit{Proof:} If $\mathbf{Z}^{sdr}$ satisfies \eqref{rankconstraint}, all of (P1)-(P4) have been solved optimally \cite{luo2010semidefinite}. In that case, because of the rank-one constraint \eqref{rankconstraint}, $\mathbf{Z}^{sdr}$ can be decomposed uniquely as $\mathbf{Z}^{sdr} = \widetilde{\mathbf{z}}^{sdr} ({\widetilde{\mathbf{z}}}^{sdr})^T$ and $\widetilde{\mathbf{z}}^{sdr}$ is optimal (feasible) for (P4). Thus, we have the following equations, 
\begin{subequations}
\begin{empheq}[left={\empheqlbrace\,}]{align}
& \mathrm{diag}(\mathbf{Z}^{sdr})_n = ({\widetilde{\mathbf{z}}^{sdr}_n})^2, \quad n = 1,2,...,I+J+2L \label{Lemma3_eq1} \\
& ({\widetilde{\mathbf{z}}^{sdr}_n})^2 = {\widetilde{\mathbf{z}}^{sdr}_n}, \quad n = 1,2,...,I+J+2L\label{Lemma3_eq2} 
\end{empheq}
\end{subequations}\eqref{Lemma3_eq1} holds because of the equation $\mathbf{Z}^{sdr} = \widetilde{\mathbf{z}}^{sdr} ({\widetilde{\mathbf{z}}}^{sdr})^T$ and \eqref{Lemma3_eq2} holds because the first $I\!+\!J\!+\!2L$ elements of $\widetilde{\mathbf{z}}^{sdr}$ can only take values from $\{0,1\}$ when it is feasible for (P4). Lemma 3 is then proven by \eqref{Lemma3_eq1}-\eqref{Lemma3_eq2} and the following two facts: (P1)-(P4) have been solved optimally if $\mathbf{Z}^{sdr}$ satisfies \eqref{rankconstraint}; the decomposition $\mathbf{Z}^{sdr} = \widetilde{\mathbf{z}}^{sdr} ({\widetilde{\mathbf{z}}}^{sdr})^T$ is unique if \eqref{rankconstraint} is satisfied. This completes the proof of Lemma 3. $\square$

Although Lemma 3 shows that the optimal solution of (P3) can be achieved by solving (P5) if $\mathbf{Z}^{sdr}$ satisfies \eqref{rankconstraint}, this rank-one constraint would not hold in most cases. Even so, we can still utilize $\mathbf{z}^{sdr}$ defined by \eqref{diagonalZ} as a start point to refine a feasible solution for (P1). To do this, we initially capture the corresponding decision vectors from $\mathbf{z}^{sdr}$, that is, 
\begin{subequations}
\begin{empheq}[left={\empheqlbrace\,}]{align}
& \mathbf{w}^{sdr} \triangleq [\mathbf{z}^{sdr}_1, \mathbf{z}^{sdr}_2,..., \mathbf{z}^{sdr}_I]^T \in \mathbb{R}^{I} \label{sdrw} \\
& \mathbf{x}^{sdr} \triangleq [\mathbf{z}^{sdr}_{I+1}, \mathbf{z}^{sdr}_{I+2},..., \mathbf{z}^{sdr}_{I+J}]^T \in \mathbb{R}^{J} \label{sdrx} 
\end{empheq}
\end{subequations}

\begin{algorithm}[!t]
\caption{SDR Heuristic Algorithm}
\label{SDRHeu}
\begin{algorithmic}[1]
\STATE Solve the problem (P5) and denote the result as $\mathbf{Z}^{sdr}$.
\STATE Obtain the $\mathbf{z}^{sdr}$, $\mathbf{w}^{sdr}$ and $\mathbf{x}^{sdr}$ by the equations \eqref{diagonalZ}, \eqref{sdrw} and \eqref{sdrx}, respectively. 
\STATE Greedily set the largest element in $\mathbf{w}^{sdr}$ to one and all other elements are set to zero. Denote the result as $\mathbf{w}^*$. \label{greedydeploy}
\REPEAT
\STATE For $j \in \mathcal{J}$, set $x_j=1$ with probability $\mathbf{x}^{sdr}_j$. Denote this feasible user association vector as $\mathbf{x}^*$. \label{randrounding}
\STATE Allocate subcarriers to users by the illustrated greedy assignment strategy. Denote the solution as $\mathbf{y}^*$. \label{greedyassign}
\STATE Update the best solution $\mathbf{x}^*$ and $\mathbf{y}^*$ so far.
\UNTIL{Reach the predefined number of iterations $t^{max}$.} \label{endrepeat} \label{stoppingcriteria}
\end{algorithmic}
\end{algorithm}

Firstly, we decide the RABS deployment greedily. Recalling that \eqref{Pro1C2} implicates that there is at most one candidate location can be chosen by RABS, that is, $\mathbf{w}$ should have at most one element equal to one and all others are zero, we greedily select the largest element in $\mathbf{w}^{sdr}$ and let it equal to one. All other elements are set to zero. Afterwards, a feasible deployment decision vector is generated and denoted by $\mathbf{w}^*$. Afterwards, a feasible user association vector $\mathbf{x}$ is constructed from $\mathbf{x}^{sdr}$ by the randomized rounding technique, i.e., set $x_j=1$ with probability $\mathbf{x}^{sdr}_j$. Denote the result as $\mathbf{x}^*$. Once the RABS deployment and user association have been fixed, the subcarriers should be allocated to users with satisfying the constraints \eqref{Pro1C3}-\eqref{Pro1C6}. By exploring the problem structure, we then illustrate a greedy assignment strategy. 

Recalling that each subcarrier can be assigned to at most one user, constrained by \eqref{Pro1C3}, and both $R^{MBS}_{jk}$ and $R^{RABS}_{ijk}$ increase monotonically with $b_k$ when the transmit power-spectral density $\zeta$ is fixed, shown by \eqref{MBSAchRate}-\eqref{RABSAchRate}, an intuitive greedy assignment strategy is then illustrated to allocate the subcarriers. Assign the available subcarrier with the largest $b_k$ to the user with the least data rate. If there are more than one users have the least rate, select the one with the greatest path loss. Repeat this process until the constraints in \eqref{Pro1C4}-\eqref{Pro1C6} are violated or there is not any unassigned subcarrier. The developed SDR heuristic method is summarized as Algorithm \ref{SDRHeu}. Note that because the randomized rounding technique is applied in line \ref{randrounding} of Algorithm \ref{SDRHeu}, we repeat the refinement procedure $t^{max}$ times and choose the best solution. 

\textit{\textbf{Remark 1:}} (Computation complexity) \cite{luo2010semidefinite} illustrates that the worst case of solving a semidefinite programming is $\mathcal{O}\big(\max\{m,n\}^4n^{1/2} \log (1/\epsilon)\big)$ with a given solution accuracy $\epsilon > 0$, where $m$ and $n$ are the number of constraints and variables, respectively. Racall that the line \ref{greedyassign} in Algorithm \ref{SDRHeu} includes a sorting algorithm and an assignment procedure, thus the worst-case complexity of the refinement strategy is $\mathcal{O}\big(t^{max}(I\!+\!J\! + \!K^2 \!+\!K\!J)\big)$. Therefore, the overall complexity of the SDR heuristic algorithm is in polynomial time. 

\section{Numerical Investigations}
\label{NumericalResults}

\begin{table}[!t]
\centering
\caption{Parameter Settings}
\label{TAB para}
\begin{tabular}{ll|ll}
\hline
Parameter & Value & Parameter & Value\\
\hline
$K$ & 20  & $b_k$ & 180 kHz \\
$\zeta$ & 1 $\mathrm{\mu}$W/Hz \cite{han2014enabling} & $b^{back}$ & 700 kHz  \\
$P^{MBS}_{max}$  & 3 W  & $P^{RABS}_{max}$ & 1 W  \\
\hline
\end{tabular}
\end{table}

In this section, numerical investigations are presented to evaluate the system performance as well as the proposed SDR heuristic algorithm.  
We consider a $1\times1 \, \text{km}^2$ area where 121 candidate locations are evenly distributed, that is, $I = 121$. The users location follow the uniform random distribution and other parameterization settings used hereafter are summarized in Table \ref{TAB para}.

\begin{figure*}[ht]
\centering
\begin{minipage}[b]{0.3\textwidth} 
\centering
\includegraphics[width=1\textwidth]{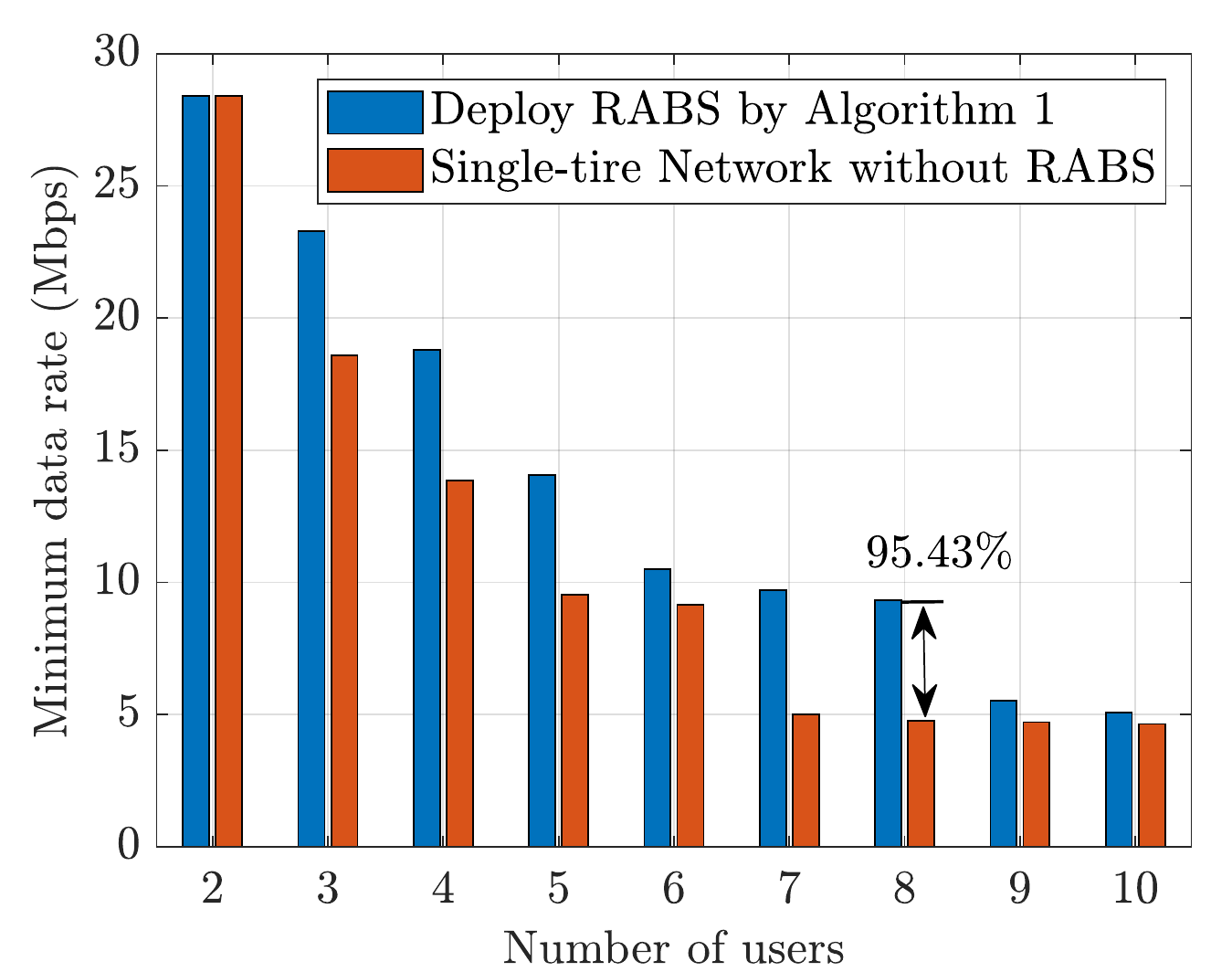} 
\caption{Minimum rate versus number of users}
\label{diffusers}
\end{minipage}
\begin{minipage}[b]{0.3\textwidth}
\centering
\includegraphics[width=1\textwidth]{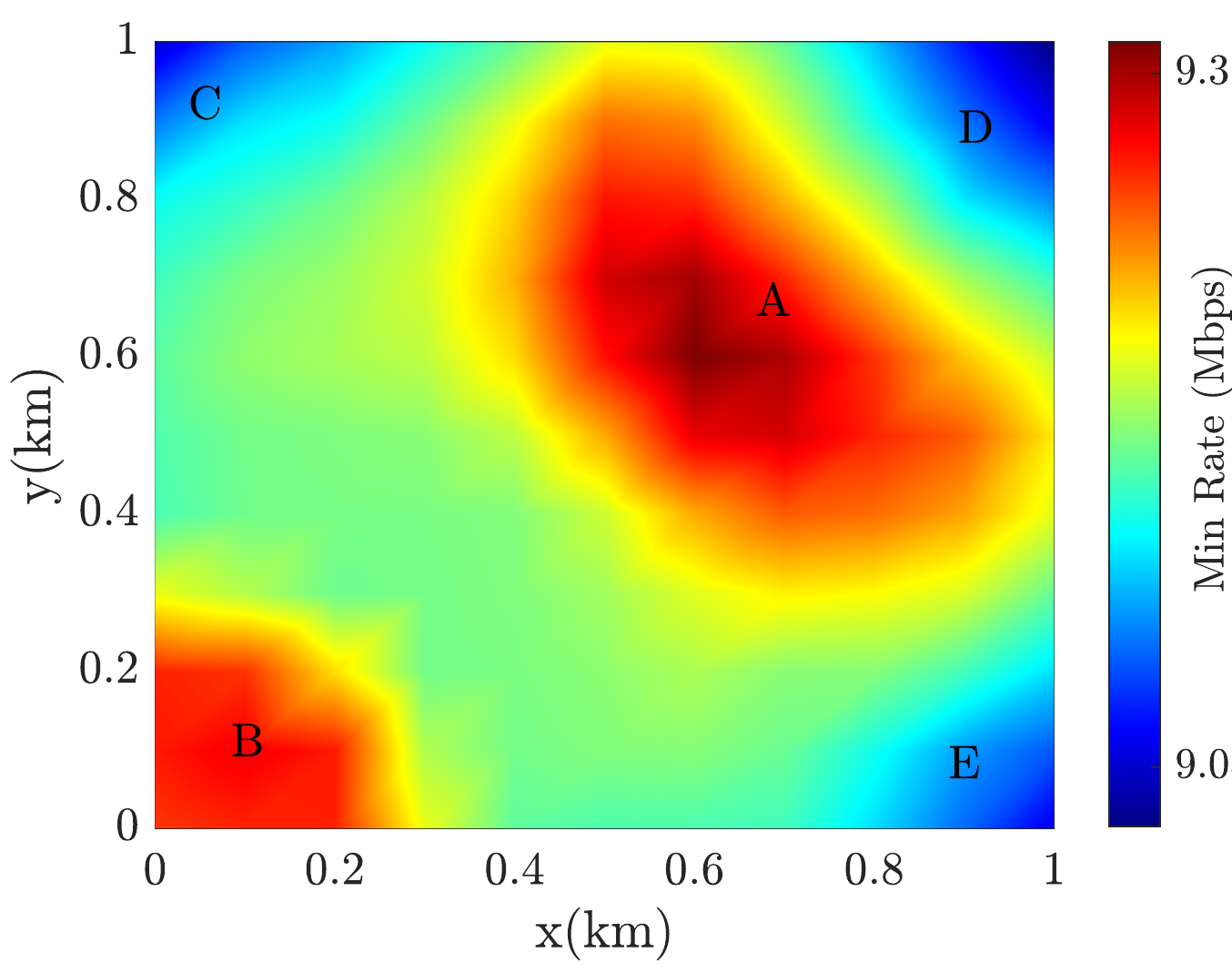}
\caption{Minimum rate versus RABS locations}
\label{RABSdep_fig}
\end{minipage}
\begin{minipage}[b]{0.3\textwidth}
\centering
\includegraphics[width=1\textwidth]{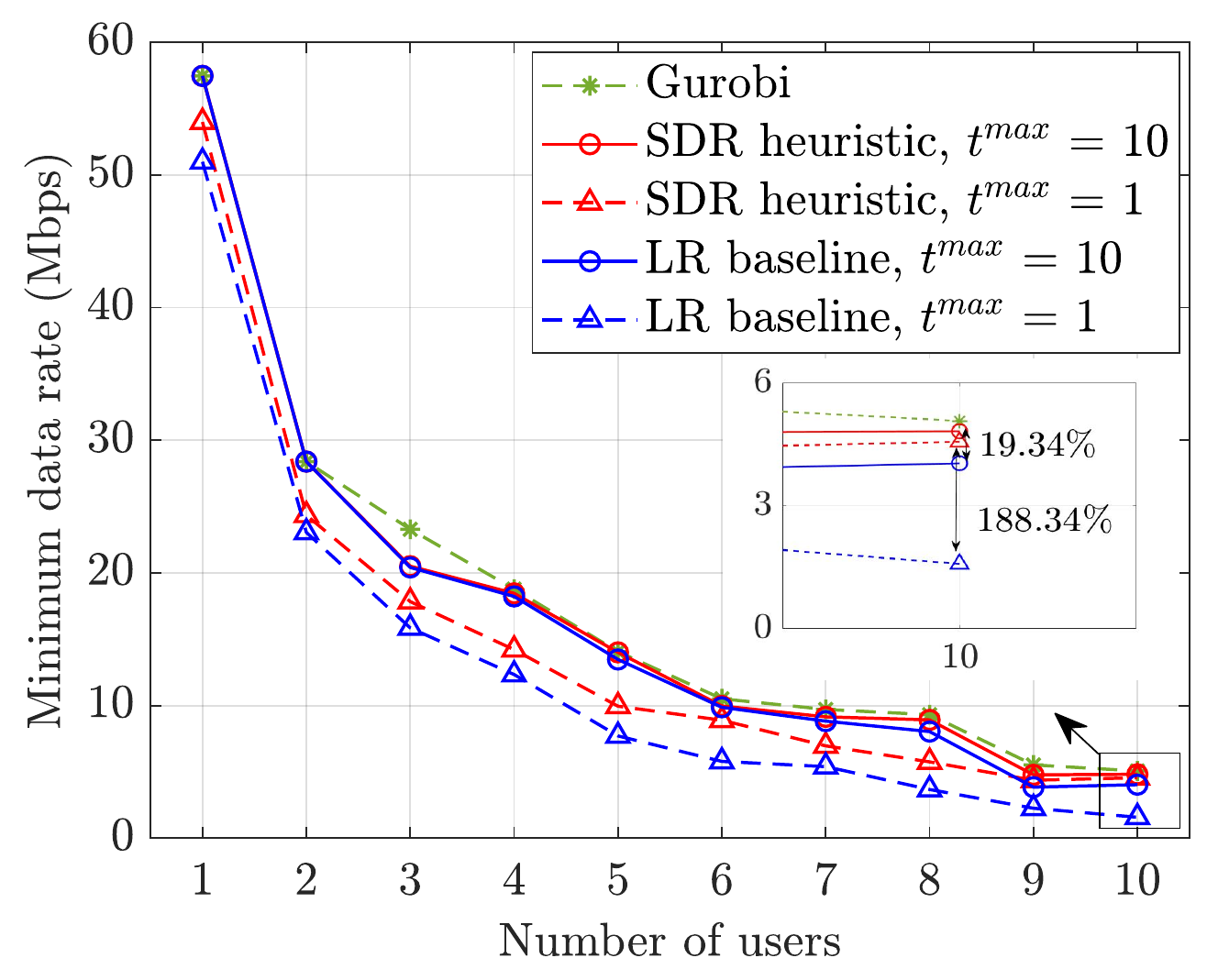}
\caption{Comparison of different algorithms}
\label{diffalg}
\end{minipage}
\end{figure*}

Fig. \ref{diffusers} compares the minimum data rate with different network configurations, i.e., a RABS-assisted HetNet versus a single-tier network with only MBS. It can be observed that deploying a RABS can improve the system performance when $I \geq 3$. However, all kinds of network have the same performance when $I \leq 2$, since in that case the minimum rate is mainly restricted by the limited capacity of backhaul link so that all users are associated to MBS. As shown in Fig. \ref{diffusers}, the performance of the RABS-assisted HetNet shows the largest gain of 95.43\% when $I = 8$ and an average gain of 33.97\%. 

Setting $I = 8$ (the point with the largest gain in Fig. \ref{diffusers}) and assuming the MBS is located at the origin point, Fig. \ref{RABSdep_fig} presents the minimum data rate versus different candidate locations via 100 Monte Carlo simulations. The best RABS deploying area, denoted by area A, is located at the centre of the macro cell edge to cover most of the cell-edge users. In contrast, when the RABS is placed at the corner areas C, D and E, the system performance is the worst because the RABS cannot cover most of the cell-edge users. For example, if the RABS is deployed at the area C, it cannot connect with the users distributed at the areas D and E, which always have low data rate because they are far from MBS. When the RABS is placed near the MBS, denoted by the area B, although it is far from the cell-edge users, the backhaul capacity increases so there is another peak appearing. Hence, region A can be the operational area of RABS, i.e., changing its perching point in lampposts that are located within that area.

The proof of Lemma 1 in Section \ref{QCQPReformulation} presents that the product of two binary variables can be linearized by introducing an auxiliary variable and a group of constraints. One might consider using the same approach to further eliminate the product term $w_i s_{jk}$ in (P2). (P2) can be then transformed to an ILP which scale grows sharply as $I$, $J$ and $K$ increase because of the combined indexes in the variables, so it is time-consuming to solve it even by state-of-art ILP solvers, such as Gurobi \cite{gurobi}. We use the linear relaxation (LR) of this ILP to replace the semidefinite programming (P5) in Algorithm \ref{SDRHeu}, implement the same refinement procedure as line \ref{greedydeploy}-\ref{endrepeat} in Algorithm \ref{SDRHeu}, and compare its performance with the proposed SDR heuristic method in Fig. \ref{diffalg}. Using the LR based method as the baseline, it can be seen that the proposed SDR heuristic method always shows a better performance. Numerically, setting $I=10$, the SDR heuristic has gains of 188.34\% and 19.34\% when setting $t^{max} = 1$ and $t^{max} = 10$, respectively. Moreover, the green dotted line in Fig. \ref{diffalg} presents the global optimal solution of (P2) solved by Gurobi \cite{gurobi}. It can be seen that the SDR heuristic with $t^{max} = 10$ has a small gap to the global optimal solution, e.g., 5.01\% when $I=10$. Notably, Remark 1 illustrates that the complexity of Algorithm \ref{SDRHeu} is in polynomial time, in contrast to the exponential worst-case complexity when using the branch and cut method to solve (P2) by Gurobi \cite{gurobi}.

\section{Conclusions}
\label{Conclusions}
Robotic aerial small cells with dexterous end effectors able to grasp at tall urban landforms can introduce significant degrees of flexibility for network densification in emerging 6G networks. This paper focuses on a HetNet consisting of a MBS and a RABS linked by a limited-capacity wireless backhaul. The minimum data rate among all users is maximized by jointly optimizing the RABS deployment location, user association and subcarrier allocation. The problem is formulated as a BPO, which is then reformulated as a QCQP and solved by a proposed SDR heuristic method in polynomial time. Numerical investigations reveal that deploying a RABS can significantly assist the network to increase the minimum achievable rate for the end users. The proposed SDR algorithm also shows a better performance than the LR baseline method. 

\bibliographystyle{IEEEtran}
\bibliography{IEEEabrv,reference}

\end{document}